# ANALYSIS OF SQUEAL NOISE AND MODE COUPLING INSTABILITIES INCLUDING DAMPING AND GYROSCOPIC EFFECTS

B. HERVÉ [1 and 2], J-J. SINOU [2 *], H. MAHÉ [1] and L. JÉZÉQUEL [2]

[1] Valeo Transmissions, Centre d'Étude des Produits Nouveaux
Espace Industriel Nord, Route de Poulainville, 80009 Amiens Cedex 1, France

[2] Laboratoire de Tribologie et Dynamique des Systèmes UMR-CNRS 5513, École Centrale de Lyon, 36 Avenue Guy de Collongue 69134 Ecully Cedex, France

## ABSTRACT

This paper deals with an audible disturbance known as automotive clutch squeal noise from the viewpoint of friction-induced mode coupling instability. Firstly, an auto-coupling model is presented showing a non-conservative circulatory effect originating from friction forces.
Secondly, the stability of an equilibrium is investigated by determining the eigenvalues of the system linearized equations. The effects of the circulatory and gyroscopic actions are examined analytically and numerically to determine their influence on the stability region. Separate and combined effects are analysed with and without structural damping and important information is obtained on the role of each parameter and their interactions regarding overall stability. Not only is structural damping shown to be of primary importance, as reported in many previous works, this article also highlights a particular relationship with gyroscopic effects.
A method of optimizing both the stability range and its robustness with respect to uncertainty on system parameters is discussed after which practical design recommendations are given.

## 1. INTRODUCTION

In cars with manual transmission, different unforced vibrations may be observed during the sliding phase of clutch engagement and are classified according to their frequency range. Such friction-induced self-generated vibrations can be explained by four general and independent mechanisms, namely stick-slip, speed dependent friction force, sprag-slip and mode coupling [1-17]. The former two rely on tribological properties whereas the latter two are due to geometrical conditions. Interested readers may refer to papers [1-4] for an overview of the four general mechanisms.

In the automotive industry, low frequency phenomena such as judder (10-20Hz), felt as jolts of the vehicle when starting, can often be attributed to tribological properties. Other works on the subject generally show good agreement between analytical predictions and experimental observations [18].

However, squeal noise is a relatively high frequency phenomenon (up to several kHz), and may arise even when the friction coefficient is almost constant versus sliding speed. It cannot be related to stick-slip behaviour because of the speed range of the vibrations measured. In this case the assumption of a purely tribological origin appears unrealistic. Consequently, sprag-slip and to a greater extent mode coupling instabilities due to the intrinsic structure of the system are more likely to be responsible for this phenomenon. We decided to focus on mode coupling for the following study because of the absence of a static sprag condition.

Firstly, a phenomenological model is presented. Then, important information is obtained concerning the influence of each parameter on the stability domain and practical design recommendations are discussed.

As reported in many previous works, structural damping is of primary importance to the stability of mode couplings [19-23]. The present study investigates in detail its influence on the effects of the combined circulatory and gyroscopic actions.

## 2. ANALYTICAL MODEL

The phenomenological model proposed is depicted in Fig. 1. It includes a homogeneous disc (A) rotating about the O**z** axis, free to swing along the O**x** and O**y** axes. O**xyz** is the principal inertial frame of (A) and is therefore a turning frame. Two sets of flexible elements in each swinging direction characterized by linearized stiffnesses $k_\theta$ and $k_\phi$ and damping $d_\theta$ and $d_\phi$ exert restoring torques on (A). The disc (A) is assumed to be in permanent contact with a second disc (B), also rotating about the O**z** axis. For the sake of simplicity, this contact is represented by four flexible elements equally distributed on radius R from axis O**z**, in planes O**xz** and O**yz** and characterized by equal stiffness $k_c$ and, for the sake of completeness, damping $d_c$. The distance between the swinging axis and the contact surface is denoted h. For the sake of simplicity, friction coefficient $\mu$ is assumed to be constant and the rotation speeds to be constant and in the same sense. The sign of the sliding speed is assumed not to change, even locally due to the fact that the observed amplitude of the vibrations remains low. Lastly, the equilibrium position is assumed to correspond to the parallelism of the two discs as illustrated in Fig. 1.

Disc (A) represents a friction disc and disc (B) the engine flywheel. They will be named the driven and driving discs respectively. The springs depicted between the two discs in the model represent the flexibility of the contact area, while h is related to the thickness of the driven disc.

This simple model is sufficient to obtain mode coupling, show different phenomena by varying stiffness and damping and propose several analytical treatments.

Writing the equations of motion linearized in the vicinity of the equilibrium for the driven disc gives

$$\mathbf{M}\ddot{\mathbf{X}} + \mathbf{D}\dot{\mathbf{X}} + \mathbf{K}\mathbf{X} = 0,  \quad (1)$$

where $\mathbf{X} = \begin{bmatrix} \theta & \phi \end{bmatrix}^T$. The mass, damping and stiffness matrices are given by

$$\mathbf{M} = \begin{bmatrix} J_\theta & 0 \\ 0 & J_\phi \end{bmatrix}, \quad (2)$$

$$\mathbf{D} = \begin{bmatrix} d_\theta + 2R^2 d_c & -(J_\theta + J_\phi - J_\varpi)\varpi - 2\mu Rh d_c.sign(\Omega - \varpi) \\ (J_\theta + J_\phi - J_\varpi)\varpi + 2\mu Rh d_c.sign(\Omega - \varpi) & d_\phi + 2R^2 d_c \end{bmatrix}, \quad (3)$$

$$\mathbf{K} = \begin{bmatrix} k_\theta + 2R^2 k_c + (J_\varpi - J_\phi)\varpi^2 & -2\mu Rh k_c.sign(\Omega - \varpi) \\ 2\mu Rh k_c.sign(\Omega - \varpi) & k_\phi + 2R^2 k_c + (J_\varpi - J_\theta)\varpi^2 \end{bmatrix}. \quad (4)$$

The upper dot denotes a derivative with respect to time t. The specific (static and uncoupled) characteristics of each DOF appear in the diagonal parts: inertia, damping and stiffness. Moreover, the rotation of the disc (A) is responsible for (diagonal) centrifugal stiffening terms and a(n) (antisymmetric skew-diagonal) gyroscopic action in the stiffness and damping matrixes respectively. Finally, the friction forces exert on each DOF a torque dependant on the position (through $k_c$) and speed ($d_c$) on the other DOF due to lever arm h, resulting in antisymmetric skew-diagonal terms, i.e. a circulatory action in the stiffness matrix and an additional gyroscopic action.

In order to generalize the associated analysis, the following non-dimensionalization is proposed, assuming $J_\theta = J_\phi = J$ for the sake of simplicity

$$\begin{bmatrix} x_1'' \\ x_2'' \end{bmatrix} + \begin{bmatrix} 2\xi & \rho \\ -\rho & 2\alpha\beta\xi \end{bmatrix} \begin{bmatrix} x_1' \\ x_2' \end{bmatrix} + \begin{bmatrix} 1 & \varphi \\ -\varphi & \alpha^2 \end{bmatrix} \begin{bmatrix} x_1 \\ x_2 \end{bmatrix} = \begin{bmatrix} 0 \\ 0 \end{bmatrix}. \quad (5)$$

The non-dimentionalization factors are given by

$$\omega_0 = \sqrt{\frac{k_\theta + 2R^2 k_c + (J_\varpi - J)\varpi^2}{J}}, \quad (6)$$

$$\xi = \frac{1}{2\omega_0} \frac{d_\theta + 2R^2 d_c}{J}, \quad (7)$$

$$\rho = -\frac{1}{\omega_0} \frac{2\mu Rh d_c.sign(\Omega - \varpi) + (2J - J_\varpi)\varpi}{J}, \quad (8)$$

$$\varphi = -\frac{1}{\omega_0^2} \frac{2\mu Rh k_c.sign(\Omega - \varpi)}{J}, \quad (9)$$

$$\alpha = \frac{1}{\omega_0}\sqrt{\frac{k_\varphi + 2R^2 k_c + (J_\varpi - J)\varpi^2}{J}}, \quad (10)$$

$$\beta = \frac{1}{2\alpha\xi\omega_0} \frac{d_\varphi + 2R^2 d_c}{J}. \quad (11)$$

The upper dash denotes a derivative with respect to the new time scale $\tau = \omega_0 t$.

Thus the system is parameterized by five values, namely the ratio of the natural pulsations α which is assumed to be higher than or equal to 1, the damping factor of the reference DOF ξ, the ratio of the damping factors β, the circulatory action φ and the gyroscopic action ρ.

## 3. STABILITY ANALYSIS

In order to investigate the stability and the influence of each parameter, an eigenvalue analysis is performed. Considering the linearized governing equation, the non-resonant solutions can be written in the complex exponential form given by

$$\mathbf{X}(\tau) = \sum a_n e^{\lambda_i(\tau + \tau_{0n})} \mathbf{X}_{0n}. \quad (12)$$

Consequently the following generalized eigenvalue equation is derived from Eq. 5.

$$\left(\lambda^2 \begin{bmatrix} 1 & 0 \\ 0 & 1 \end{bmatrix} + \lambda \begin{bmatrix} 2\xi & \rho \\ -\rho & 2\alpha\beta\xi \end{bmatrix} + \begin{bmatrix} 1 & \varphi \\ -\varphi & \alpha^2 \end{bmatrix}\right) \begin{bmatrix} x_{10} \\ x_{20} \end{bmatrix} = \begin{bmatrix} 0 \\ 0 \end{bmatrix}. \quad (13)$$

Eigenvalues and associated eigenvectors are obtained by solving this equation. The eigenvectors represent the mode shape. The imaginary part of the associated eigenvalues is the pulsation and the real part is the exponential growth rate of the amplitude of the modes. According to the Hartman-Grobman theorem, when the real parts of the eigenvalues of the linearized system are negative, the equilibrium studied is stable and when at least one is positive, the equilibrium then becomes unstable independently of the non-linear behaviour.

## 4. RESULTS WITHOUT STRUCTURAL DAMPING

In this section, the effects of the circulatory and gyroscopic actions on the eigenelements and the stability of the equilibrium are investigated by assuming no structural damping.

### 4.1. Purely circulatory system

The corresponding equation is given by

$$\left(\lambda^2 \begin{bmatrix} 1 & 0 \\ 0 & 1 \end{bmatrix} + \begin{bmatrix} 1 & \varphi \\ -\varphi & \alpha^2 \end{bmatrix}\right) \begin{bmatrix} x_{10} \\ x_{20} \end{bmatrix} = \begin{bmatrix} 0 \\ 0 \end{bmatrix}. \quad (14)$$

The eigenvalues and eigenvectors are derived. We obtain

$$\lambda^2 = -\frac{\alpha^2 + 1}{2} \pm \sqrt{\left(\frac{\alpha^2 - 1}{2}\right)^2 - \varphi^2} \quad (15)$$

and

$$\left[\begin{pmatrix} x_{10} \\ x_{20} \end{pmatrix}_{n=1..4}\right] = \begin{bmatrix} \varphi \\ -(\lambda_n^2 + 1) \end{bmatrix}. \quad (16)$$

Four eigenvalues and only two linearly independent modal vectors are obtained in this circumstance, related to the eigenelements of the sole stiffness matrix, as suggested by Eq. 14. From Eq. 15, a critical value can be derived for the amplitude of φ

$$|\varphi| = \left|\frac{\alpha^2 - 1}{2}\right|. \quad (17)$$

Below this limit, the eigenvalues are purely imaginary thus, according to Eq. 16, the eigenvectors have a real ratio between their two components, thus the modal trajectories are non-turning trajectories. As φ increases, the two pulsations approach each other and the direction of the two modes tends to be the same until the modes merge at the coalescence point. As soon as the limit value is exceeded the squared eigenvalues become conjugate complex, thus the eigenvalues display two opposite pairs of conjugate values. As a consequence the system has a unique pulsation and the equilibrium is unstable due to the eigenvalues with a positive real part, leading to flutter. This is the basic destabilizing effect due to the circulatory action.

Moreover, the modal trajectories become turning ones and turn in opposite senses since the eigenvectors are conjugate and, as shown in Eq. 16, the modal trajectory associated with the unstable eigenvalues turns in the direct sense when the circulatory action is positive and vice-versa.

Direct and reverse senses are relative to oriented variable space $x_1$, $x_2$ and is different from the usual sense taken into account when working on rotating machinery, which is relative to rotation.

As an example, a negative circulatory action is obtained in Eq. 4 when $\Omega > \varpi$. In this situation, for a positive value of θ the friction forces induce a torque that tends to decrease ϕ; for a negative value of ϕ, it tends to decrease θ. This is the basic antisymmetric coupling mechanism of the system. Thus, the associated motion appears to be turning in reverse sense, in accordance with the previous conclusion and the physical intuition according to which the rotation of the unstable modal trajectory is oriented by the circulatory action.

A typical evolution of the eigenvalues is depicted in Fig. 2 and the associated modal trajectories are in the centre line of Fig. 3, corresponding to a null gyroscopic action.

### 4.2. Purely gyroscopic system

The following equation of the purely gyroscopic undamped system is considered

$$\left(\lambda^2 \begin{bmatrix} 1 & 0 \\ 0 & 1 \end{bmatrix} + \lambda \begin{bmatrix} 0 & \rho \\ -\rho & 0 \end{bmatrix} + \begin{bmatrix} 1 & 0 \\ 0 & \alpha^2 \end{bmatrix}\right) \begin{bmatrix} \theta_0 \\ \varphi_0 \end{bmatrix} = \begin{bmatrix} 0 \\ 0 \end{bmatrix}. \tag{18}$$

The characteristic equation and its roots are then given by

$$\lambda^4 + (1 + \alpha^2 + \rho^2)\lambda^2 + \alpha^2 = 0 \tag{19}$$

and

$$\lambda^2 = -\frac{1 + \alpha^2 + \rho^2}{2} \pm \sqrt{\left(\frac{1 + \alpha^2 + \rho^2}{2}\right)^2 - \alpha^2}. \tag{20}$$

The term in the square root is positive, thus the square root is real; moreover this root has a smaller absolute value than the preceding term. Thus the eigenvalues remain purely imaginary and, as the gyroscopic action increases, the highest pulsation increases to infinite while the lowest pulsation decreases to zero, producing the well-known Campbell effect.

In this system, four modal vectors can be written in the form

$$\left[\begin{pmatrix} x_{10} \\ x_{20} \end{pmatrix}_{n=1..4}\right] = \begin{bmatrix} \rho \lambda_n \\ -(\lambda_n^2 + 1) \end{bmatrix}. \tag{21}$$

This result shows the gyroscopic splitting of the modes. The four vectors are complex, and the modal trajectory associated with each squared eigenvalue rotates in the opposite senses. The sign of the gyroscopic action does not influence the pulsations, but determines the sense of rotation of the associated modal trajectory. The modal trajectory turning in the direct sense is associated with the highest pulsation when the gyroscopic action is positive and vice-versa.

It can be concluded that the gyroscopic action alone changes the pulsations and transforms the modal trajectories into turning trajectories but does not destabilize the equilibrium. The modal trajectories are in the central column of Fig. 3, corresponding to a null circulatory action.

### 4.3. Combination of circulatory and gyroscopic actions

Finally, in presence of both gyroscopic and circulatory actions, the modal equation becomes

$$\left(\lambda^2 \begin{bmatrix} 1 & 0 \\ 0 & 1 \end{bmatrix} + \lambda \begin{bmatrix} 0 & \rho \\ -\rho & 0 \end{bmatrix} + \begin{bmatrix} 1 & \varphi \\ -\varphi & \alpha^2 \end{bmatrix}\right) \begin{bmatrix} x_{10} \\ x_{20} \end{bmatrix} = \begin{bmatrix} 0 \\ 0 \end{bmatrix}. \tag{22}$$

The associated characteristic equation is given by

$$(\lambda^2 + 1)(\lambda^2 + \alpha^2) + (\rho\lambda + \varphi)^2 = 0. \tag{23}$$

The equation cannot be balanced with either an assumed purely imaginary or a real value. Therefore the eigenvalues always have both non-zero imaginary and real parts and are conjugated by pair since the characteristic equation is real. Moreover, since the third order term of the characteristic polynomial is null, the two pairs of eigenvalues have opposite real parts.

As a consequence, the equilibrium of the undamped system subjected to a combined circulatory and gyroscopic action is systematically unstable and leads to flutter.

Moreover, by using Eq. 23 it is possible to check that the proportion factor between the two modal vectors expressed in Eq. 24 is not real, except in the case of the purely circulatory system below the coalescence point.

$$\left[\begin{pmatrix} x_{10} \\ x_{20} \end{pmatrix}_{n=1..4}\right] = \begin{bmatrix} \rho \lambda_n + \varphi \\ -(\lambda_n^2 + 1) \end{bmatrix}. \quad (24)$$

Thus, because of the continuity of the roots of a polynomial with respect to its coefficients, the sense of rotation of the modal trajectory associated with each individual eigenvalue remains the same when traversing the parameter space. However in the case of the purely circulatory system below the coalescence point, it has been shown that the modal trajectories do not turn.

Therefore, the sense of rotation of the unstable and stable modal trajectories remains determined by the sign of the circulatory action, and the relationship between the corresponding pulsations remains determined by the sign of the gyroscopic action.

Fig. 4 shows the pulsations and the associated exponential growth rate in the gyroscopic/circulatory plane. Bold lines are also plotted to indicate the locus of the intersection and junctions. Fig. 3 shows the associated modal trajectories drawn at a few points of the plane. These figures provide a complete overview of the qualitative behaviour of the undamped system regarding the stability of the equilibrium. Indeed, the results given in all the previous discussions do not depend qualitatively on the ratio of the natural pulsations or the domain of study, assuming it includes the coalescence point.

The region of potential stability (limiting case of the Hartman-Grobman theorem) can be observed as corresponding either to a purely gyroscopic or a purely circulatory action (within a limited range), therefore the robustness of the stability is null (its sensitivity is infinite) with respect to the vanishing action.

From this point, the fact that neither the gyroscopic nor the circulatory action could be neglected *a priori* at the same time as damping in a stability investigation should be strongly underlined.

## 5. RESULTS WITH STRUCTURAL DAMPING

In the presence of strictly positive structural damping the characteristic polynomial shows a strictly negative third order coefficient, which indicates that the sum of the real parts of the (paired conjugate) eigenvalues is negative. As a consequence only one pulsation at the same time can be associated with an unstable nature.

The stable and unstable domains are separated by frontiers due to the continuity of the eigenvalues. Assuming one pair of eigenvalues has a null real part and identifying the coefficients of the characteristic polynomial, the following equation of the frontiers is derived

$$(\varphi\rho + \alpha\xi(\alpha+\beta))^2 - \xi(1+\alpha\beta)(\varphi\rho+\alpha\xi(\alpha+\beta))(1+\alpha^2+\rho^2+4\alpha\beta\xi^2) + \xi^2(1+\alpha\beta)^2(\alpha^2+\varphi^2) = 0. \quad (25)$$

This is a second order polynomial of φ and a third order polynomial of ρ. When the gyroscopic action tends to infinite, the circulatory actions on the frontier of stability tend to the two following equations respectively,

$$\varphi = 0, \quad (26)$$

$$\varphi = \rho\xi(1+\alpha\beta) \quad (27)$$

describing two asymptotic lines for the frontier.

Moreover, it can be shown that the frontiers remain outside the asymptotes when no DOF is overdamped. The asymptotes can therefore be considered as stability frontiers by default.

In order to clarify the global influence of the damping matrix comparatively to the basic destabilizing effect of the circulatory action, we propose writing the equation of motion in the eigenbase of the stiffness matrix. Assuming a non-null circulatory action different from the coalescence value, the eigenvalues and the associated eigenvectors of the stiffness matrix are given by

$$\begin{cases} e_1 = \dfrac{\alpha^2 + 1}{2} - \sqrt{\left(\dfrac{\alpha^2 - 1}{2}\right)^2 - \varphi^2} \\ e_2 = \dfrac{\alpha^2 + 1}{2} + \sqrt{\left(\dfrac{\alpha^2 - 1}{2}\right)^2 - \varphi^2} \end{cases} \quad (28)$$

and

$$[\boldsymbol{\psi}_1 \quad \boldsymbol{\psi}_2] = \begin{bmatrix} \alpha^2 - e_1 & \varphi \\ \varphi & e_2 - 1 \end{bmatrix}. \quad (29)$$

In the new base the equation of motion can be re-written as in Eq. 30 with the new matrix expressed in Eq. 31-33.

$$\boldsymbol{\Psi}'' + \boldsymbol{\Delta}\boldsymbol{\Psi}' + \boldsymbol{\Gamma}\boldsymbol{\Psi}' + \mathbf{K}\boldsymbol{\Psi} = \mathbf{0}, \quad (30)$$

$$\mathbf{K} = \begin{bmatrix} e_1 & 0 \\ 0 & e_2 \end{bmatrix}, \quad (31)$$

$$\boldsymbol{\Delta} = \left[\left(\dfrac{\alpha^2 - 1}{2}\right)^2 - \varphi^2\right]^{-\frac{1}{2}} \left((\alpha^2 - 1)\xi\begin{bmatrix} 1 & 0 \\ 0 & \alpha\beta \end{bmatrix} + \varphi\rho\begin{bmatrix} 1 & 0 \\ 0 & -1 \end{bmatrix} - \dfrac{\varphi^2 \xi(\alpha\beta + 1)}{e_2 - 1}\begin{bmatrix} 1 & 0 \\ 0 & 1 \end{bmatrix}\right), \quad (32)$$

$$\boldsymbol{\Gamma} = \left[\left(\dfrac{\alpha^2 - 1}{2}\right)^2 - \varphi^2\right]^{-\frac{1}{2}} \left(\left(\dfrac{\alpha^2 - 1}{2}\right)\rho - (\alpha\beta - 1)\xi\varphi\right)\begin{bmatrix} 0 & 1 \\ -1 & 0 \end{bmatrix}. \quad (33)$$

In this expression the new stiffness matrix is intentionally diagonal and all the coupling terms have been combined in the damping matrix, split into a diagonal damping element $\boldsymbol{\Delta}$ and a purely gyroscopic element $\boldsymbol{\Gamma}$. It should be noted that the "total" gyroscopic action obtained not only includes the "external" gyroscopic action but also a contribution from the circulatory action through the defect of proportionality of the structural damping.

While the circulatory action is below the coalescence point, the base vectors are real and all the terms are real. As a consequence the system can be investigated as a purely gyroscopic real system. When the circulatory action is beyond the coalescence point, the base vectors are complex as are the terms. In this case, the system cannot be investigated as a purely gyroscopic real system. Nevertheless some characteristic behaviour remains unchanged.

Fig. 5 shows a characteristic plot of the frontiers in the (original) gyroscopic/circulatory action plane with and without the total gyroscopic action. Below the coalescence point, the Eq. 30 is real. Without considering $\boldsymbol{\Gamma}$ the diagonal system shows unstable equilibrium only when one of the diagonal damping terms of $\boldsymbol{\Delta}$ becomes negative. Beyond the coalescence point, the criterion becomes slightly more complicated since the system becomes complex. Whatever the case, the frontiers clearly appear to be different all over the gyroscopic/circulatory action plane with and without $\boldsymbol{\Gamma}$.

This illustration shows that the intuitive idea according to which the stability of equilibrium would only be related to the terms of $\boldsymbol{\Delta}$ is false. Indeed, the only coinciding point corresponds to the cancellations of $\boldsymbol{\Gamma}$. Thus it appears that the terms of both $\boldsymbol{\Delta}$ and $\boldsymbol{\Gamma}$ both influence stability. In order to further understand the role played by the different elements in stability, we provide results on the purely gyroscopic and purely circulatory damped systems in the following parts which include the examination of the complete system.

## 5.1. Purely gyroscopic system

Without circulatory action, the equation of motion fits the desired form naturally

$$\Psi'' + \lambda \begin{bmatrix} 2\xi & \rho \\ -\rho & 2\alpha\beta\xi \end{bmatrix} \Psi' + \begin{bmatrix} 1 & 0 \\ 0 & \alpha^2 \end{bmatrix} \Psi = \begin{bmatrix} 0 \\ 0 \end{bmatrix}. \tag{34}$$

Thus Eq. 25 can be re-written giving

$$\beta(\alpha+1)^2(\alpha-1)^2 + (1+\alpha\beta)(\alpha+\beta)(\rho^2 + 4\alpha\beta\xi^2) = 0. \tag{35}$$

No real value of ρ can be obtained from this expression. What is more, since the sum of the real part of the eigenvalues is negative, the equilibrium is always asymptotically stable.

In order to investigate the sense of rotation of the modal trajectories, a modal base can be written in the original base

$$\left[ \begin{pmatrix} x_{10} \\ x_{20} \end{pmatrix}_{n=1..4} \right] = \begin{bmatrix} \rho\lambda_n \\ -(\lambda_n^2 + 2\xi\lambda_n + 1) \end{bmatrix}. \tag{36}$$

This expression and the characteristic polynomial permit checking that that the relationship between the pulsations and the sense of rotation is the same as in the undamped case.

The modal trajectories are illustrated in Fig. 6. The asymptotic stability as well as the association of the pulsations to the senses of rotation can be noted. This should be compared to the central column of Fig. 3 corresponding to the associated undamped system.

According to Eq. 30-33, the behaviour of the complete system below the coalescence point can be directly linked to these results. Moreover, the latter allow explaining the influence of **Γ** on the stability frontiers illustrated in Fig. 5. Writing Eq. 30, which is assumed to be real, in the form of Eq. 34 with not necessarily positive diagonal damping terms, with ξ and β being either positive or negative, gives Eq. 35 for the frontiers. In this case, the existence of real (opposing) frontiers requires β to be negative. The stability at the origin is obtained only when both the diagonal damping terms are positive or null and vice-versa, which allows determining the stable and unstable sides of the frontiers when they exist. When β is null, the equilibrium is stable at the origin and asymptotically stable for any non-null value of ρ.

These few observations are sufficient to show that the frontiers obtained below the coalescence point, without considering **Γ** in Eq 30, are stability frontiers by default, as illustrated in Fig. 5.

The analysis of the system's behaviour beyond the coalescence point could also be performed in a comparable manner but would require taking into consideration complex structural damping and gyroscopic action.

## 5.2. Purely circulatory system

The stability analysis for the purely circulatory damped system is of particular interest when the gyroscopic action is slight. In this part, the stability in the case of the system described by

$$\left( \lambda^2 \begin{bmatrix} 1 & 0 \\ 0 & 1 \end{bmatrix} + \lambda \begin{bmatrix} 2\xi & 0 \\ 0 & 2\alpha\beta\xi \end{bmatrix} + \begin{bmatrix} 1 & \varphi \\ -\varphi & \alpha^2 \end{bmatrix} \right) \begin{bmatrix} x_{10} \\ x_{20} \end{bmatrix} = \begin{bmatrix} 0 \\ 0 \end{bmatrix} \tag{37}$$

is investigated. The following equation for the frontiers can then be derived from Eq. 25.

$$\varphi^2 = \frac{\alpha\beta}{(1+\alpha\beta)^2} \left[ (\alpha^2-1)^2 + 4\alpha\xi^2(1+\alpha\beta)(\alpha+\beta) \right]. \tag{38}$$

Since the damping parameter ξ is strictly positive, the equilibrium is stable at the origin and two opposite real values of the circulatory action delimit the stability domain, in a comparable manner to the undamped case. However, it should be noted that the stability domain is not necessarily broader and even tends to the sole origin as the damping ratio β tends to zero.
Therefore both the amount of damping and damping distribution have a strong influence on stability. Varying β alone not only changes the damping ratio between the two DOF but also modifies the amount of damping of the system. Even if a possible increase of stability with the total amount of damping seems favourable, it should be underlined that such unbalanced evolution may lead to under-utilization of its stabilizing capacity. In order to distinguish the respective effects of the amount and distribution of damping over the two DOF, the total damping expressed

$$\xi_t = \xi(1+\alpha\beta) \tag{39}$$

is considered, which leads to the new expression

$$\varphi^2 = \frac{\alpha\beta}{(1+\alpha\beta)^2}\left[(\alpha^2-1)^2 + 4\alpha\xi_t^2\frac{\alpha+\beta}{1+\alpha\beta}\right] \tag{40}$$

for the stability limits. Assuming the damping is low enough to preserve the order of the pulsations, it can be seen that the width of the stability domain increases with the ratio of the natural pulsations and with the amount of damping.
Nevertheless, regarding damping distribution, its evolution is not monotonic and always shows one unique maximum at the optimum value given by Eq. 41. The existence of this optimal distribution allows concluding that damping may have a destabilizing effect. Moreover, at this optimum, the stability domain has null sensitivity, or infinite robustness with respect to the sole damping distribution.

$$\beta_{opt} = \frac{\sqrt{16(\alpha^4-\alpha^2+1)\xi_t^4 + 4(\alpha^2+1)(\alpha^2-1)^2\xi_t^2 + (\alpha^2-1)^4} - 4(\alpha^2-1)\xi_t^2}{\alpha(4\xi_t^2 + (\alpha^2-1)^2)}. \tag{41}$$

As the damping amount tends to zero, the value defined in Eq. 41 tends to 1/α, which corresponds to a proportional distribution. Both criteria are compared in Fig. 7 in the damping amount/natural pulsations ratio plane. For a total damping amount up to 20%, the difference between the exact and approximated values is small. Therefore, the proportional distribution can be assimilated with the optimum for the low damping purely circulatory system.
In order to explain this optimum, it should be noted that it corresponds to **Γ** being null in Eq. 30. Thus the system behaves qualitatively as being without any coupling action. Below the coalescence point and for low damping, the pulsations remain very close to the undamped case and the real part is common to the two eigenvalues, since the terms of **Δ** are equal on each DOF, as indicated by Eq. 32. The eigenvalues are equal at the coalescence point, but beyond it the real parts become separated whereas the pulsations remain equal.
The similarities to the undamped case are clearly illustrated in Fig. 8 which shows the evolution of the eigenvalues with respect to circulatory action for various damping distributions. The evolution of the real parts, shown in Fig. 8b, highlights the correspondence of the widest range of stability to a configuration close to the proportional damping distribution.
Outside the proportional damping case, the eigenvalues are observed to have a splitting effect that can be attributed to the additional gyroscopic term related to the damping distribution in Eq. 33. However, from the coalescence point it is still possible to observe a rapid separation of the real parts and a parallel trend for the pulsations. This indicates that the coalescence phenomenon still occurs and is only partially blurred by the additional gyroscopic action.
In order to investigate the sense of rotation of the modal trajectories, the following modal base can be proposed

$$\left[\begin{pmatrix} x_{10} \\ x_{20} \end{pmatrix}_{n=1..4}\right] = \left[\begin{matrix} \varphi \\ -\left(\lambda_n^2 + 2\xi\lambda_n + 1\right) \end{matrix}\right]. \qquad (42)$$

From this expression and the characteristic polynomial, it can be shown that non-turning modal trajectories exist only for the proportional distribution and below the coalescence point. Therefore, the optimised stability also corresponds to the existence of non-turning modal trajectories.

Moreover, the unstable eigenvalue is continuously the eigenvalue having the highest real part at the origin, i.e. the lowest diagonal damping term in Eq. 32 for φ=0, as illustrated on Fig. 9. Nevertheless, the sense of rotation of the unstable modal trajectory remains related to the sign of the circulatory action, and the relationship between the pulsation distribution and the senses of rotation with the sign of the total gyroscopic action.

These observations generalize the undamped case, considering the total gyroscopic action instead of the sole "external" gyroscopic action. This is illustrated in Fig. 10 which shows the modal trajectories versus the circulatory action and for various damping distributions.

### 5.3. Stability analysis of the complete system

Damping has two effects on the system. Firstly, the presence of positive damping globally decreases the real parts of the eigenvalues since their sum becomes negative, which is the stabilizing effect expected. Secondly, the damping distribution modifies the total gyroscopic action so that the topology of the surfaces may be strongly modified.

This latter effect is illustrated in Figs. 11 and 12 which show the eigensurfaces for the proportionally damped and unbalanced damped systems respectively. Both of them have the same total damping amount and the same natural pulsation ratio, as for Fig. 4. It appears that (low) damping has a rather weak effect on pulsations, Figs. 4a, 11a and 12a, whereas the real part of the eigenvalues is strongly affected, Figs. 4b, 14b and 12b.

In comparison to the associated undamped system, the proportionally damped system shows a very slight deformation of the surfaces of the real parts, so that they remain topologically similar. The destabilizing separation of the real parts still occurs beyond the coalescence point for the purely circulatory system, and when both circulatory and gyroscopic actions are combined as in the undamped case. However, this destabilizing separation is partially counterbalanced by the generally lower amount of the real parts so that the domain of potential stability in the undamped case becomes a domain of asymptotic stability in the damped case and the separation of the real parts no longer leads to immediate destabilization.

In the unbalanced case, the topology of the surfaces of the real parts is clearly different; they are separated at the origin and their junction curve is discontinuous. Moreover, for the pulsation surfaces, the junction curves no longer correspond to a null gyroscopic action. Thus the purely circulatory system shows a pseudo-coalescence phenomenon with neither perfect merging of the pulsations nor equality of the real parts as shown previously.

These modifications also affect the stability frontiers; these are shown in Figs. 13 and 14 for different values of the total damping amount and different distributions respectively.

The evolution of the frontiers in the proportional configuration with respect to the damping amount is in continuity with the undamped case. The frontiers smoothly envelop the frontiers of the associated undamped system, highlighting the antagonistic effect of damping and destabilization due to the combination of the actions.

As the damping distribution evolves, it can be seen that the frontiers deform progressively, in such a way that the bump associated with the coalescence phenomenon does not stay on the circulatory action axis and is deviated towards a direction combining both actions. At the same time, the frontiers contract in the vicinity of the origin, up to complete collapse when only one DOF is damped. This is the cause of the previously observed restriction of the stability domain to the sole origin in the purely circulatory system.

Finally, the additional gyroscopic term also affects the relationship between the pulsation distribution and the sense of rotation of the modal trajectories. In the undamped and proportionally damped cases, not-turning modal trajectories exist only for the purely circulatory system. In the presence of gyroscopic action, the modal trajectories are turning ones with a sense of rotation related to the pulsation distribution in accordance with the sign of the gyroscopic action. This is illustrated in Fig. 15 which shows the modal trajectories at several points of the circulatory/gyroscopic action plane and the locus of existence of non-turning modal trajectories for the proportionally damped system.

However, for the unbalanced damped systems and for a circulatory action below the coalescence point, the effect of the additional gyroscopic term is in agreement with the intuitive idea that the sense of rotation of the modal trajectories, as well as the existence of non-turning trajectories, is not related only to the external gyroscopic action but to the total gyroscopic action. The modal trajectories are illustrated in Fig. 16. The locus of existence of non-turning modal trajectories appears to be the segment where the total gyroscopic action is null below the coalescence point. An interpretation using Caughey's condition also gives the same locus for the real modes. Moreover, it can be observed that this segment nearly connects with the top of the bump of the frontiers.

### 5.4. Optimization of low damping quasi-purely circulatory systems

Observation of the behaviour of the frontiers with respect to damping makes it possible to propose a stability optimisation strategy for the quasi-purely circulatory systems, as a generalization of the purely circulatory case. This also requires knowledge of the relationship between the gyroscopic and circulatory actions. Two major aspects have to be taken into consideration: firstly the width of the stability region, and secondly the robustness of the criterion with respect to the uncertainties on parameters, especially the gyroscopic actions.

In a stability investigation, gyroscopic action may be considered from two different angles: either it is assumed to be "external" and constant, as the consequence of a given rotation speed, or it is assumed to be linked to the circulatory action, as the gyroscopic action induced by the friction forces in Eq. 3. The load exerted on the system therefore shows different trajectories in the gyroscopic/circulatory action plane that depend on this consideration.

Assuming a low constant gyroscopic action, one way of optimizing the stability range with respect to the circulatory action is to set the damping distribution so that the bumps associated with the coalescence phenomenon are located at the desired value of the gyroscopic action. Generally, assuming a relationship between the two actions provides a similar optimization strategy. Nevertheless, these apply only for low gyroscopic actions since it appears that the influence of the damping distribution on the location of the bumps is restricted.

In the particular case of a linear relationship between circulatory and gyroscopic actions, for low damping systems it can be seen that the optimized system is close to that displaying non-turning modal trajectories in the stable area, similar to the undamped case. This particular property is a generalization of the criterion for systems having a null total gyroscopic action.

Regarding robustness, the effect of proportional damping is clearly positive compared to the undamped system, in both cases of purely circulatory and gyroscopic actions. However, when investigating a system combining the two actions, the criteria of the sole damping distribution for optimizing the stability range and the robustness may be antagonistic. An example of this is when the stability optimizations lead to placing most of the damping on only one DOF, the robustness of the system for a null action tends to zero with respect to a purely circulatory action. Therefore, optimizing both stability and robustness may require acting on both the damping amount and the damping distribution.

## 6. PRACTICAL DESIGN FOR AVOIDING SQUEAL NOISE

In the proposed model, Fig. 1 and Eq. 1-4, the circulatory action controlled by the friction coefficient is chosen as the control parameter. Without the gyroscopic action originating from the rotation, the curve of the parameterized load in the gyroscopic/circulatory action plane is a straight line crossing the origin. The rotation of the disc provides an additional constant gyroscopic action and centrifuge stiffening. The additional constant gyroscopic term then acts as an offset on the load line, whereas the centrifuge stiffening modifies both diagonal stiffness terms, and therefore the reference natural pulsation and natural pulsation ratio. However, in the pulsation range of squeal noises, these latter effects can be neglected.

Both these gyroscopic actions and the friction forces responsible for the circulatory action in the proposed model are related to the rotations. Therefore, assuming a constant ratio of natural pulsations, it appears possible to apply certain optimization strategies to improve the stability domain and robustness in order to reduce the propensity of clutches to squeal.

The most influential parameter allowing stabilization is the ratio of the natural pulsations. Indeed, by changing the inertia and stiffness associated to the coupled modes, the coalescence point may be shifted, thus stabilizing the equilibrium. The possible gyroscopic effects as well as their relationship with the friction forces then have to be evaluated in order to choose an optimized stabilization strategy.

## 7. CONCLUSION

A linearized two DOF mode-coupling model has been studied from the viewpoint of stability with respect to its parameters. Particular attention has been paid to damping distribution and its gyroscopic aspect, both analytically and through numerical illustrations. The results on the determination of stability frontiers are of particular interest in the case of quasi-purely circulatory mechanisms such as clutch squeal noise.

Provided the damping is proportional it has been shown that the stability range of the quasi-purely circulatory damped systems could be approached via their associated purely circulatory undamped systems. Indeed, in this configuration, the effect of damping is to add robustness to the stability domain rather than to change its range or shape. Therefore, neglecting slight gyroscopic actions is possible and the coalescence point remains close to Hopf's bifurcation point. Moreover, the modal trajectories remain the same.

However, in the general case of an unspecified damping distribution, such an approach becomes irrelevant. Because the external gyroscopic action is likely to interact with the gyroscopic aspect of damping, it can no longer be neglected. Moreover, both the stability domain and its robustness become strongly related to the load curve in the circulatory/gyroscopic action plane.

As a consequence, damping is a key parameter which requires highly detailed analysis when modelling the occurrence of instabilities. Design can be optimized from the viewpoint of stability and robustness by efficient control of the damping structure of the system relative to circulatory and gyroscopic actions.

## NOMENCLATURE

| | |
|---|---|
| $Oxyz$ | Principal inertia frame of the driven disc |
| $R$ | Radius of the contact locus between the driven and the driving discs |
| $h$ | Distance between the driven and driving discs |
| $\theta, \varphi$ | Swinging angle of the driven disc about the Ox and Oy axis respectively |
| $J_\varpi, J_\theta, J_\phi$ | Inertia of the driven disc about the Oz, Ox and Oy axis respectively |
| $\varpi, \Omega$ | Rotation speed of the driven and driving discs respectively about the Oz axis |
| $k_\theta, d_\theta, k_\phi, d_\phi$ | Swinging stiffness and damping about the Ox and Oy axis respectively |
| $k_c, d_c$ | Stiffness and damping respectively about the Oz axis at each contact points |
| $\mu$ | Friction coefficient |
| $x_1, x_2, \psi_1, \psi_2$ | Variables coordinates |
| $\mathbf{X}, \mathbf{\Psi}$ | Variable vectors |
| $x_{10}, x_{20}$ | Coordinates of eigenvectors |
| $\mathbf{X}_{0n}, a_n, \tau_{0n}$ | Vector, contribution factor and time origin respectively of the n[th] mode |
| $\mathbf{M}, \mathbf{D}, \mathbf{K}$ | Mass, damping and stiffness matrixes respectively |
| $\omega_0, \tau$ | Natural pulsation and period respectively of the reference DOF |
| $\alpha$ | Ratio of the natural pulsations of the two DOF |
| $\xi, \xi_t$ | Reference damping, total damping respectively |
| $\beta, \beta_{opt}$ | Ratio of damping and optimised ratio of damping respectively |
| $\varphi, \rho$ | Amplitude of the circulatory and gyroscopic actions respectively |
| $\bullet, \ddot{\bullet}, \bullet', \bullet''$ | Single and double derivative with respect to time t and $\tau$ respectively |
| $\lambda$ | Complex eigenvalue |
| $e_1, e_2$ | Eigenvalues of the non-dimensionalised stiffness matrix |
| $\mathbf{\Delta}, \mathbf{\Gamma}, \mathbf{K}$ | Diagonal and skew-diagonal damping, and stiffness matrixes respectively |

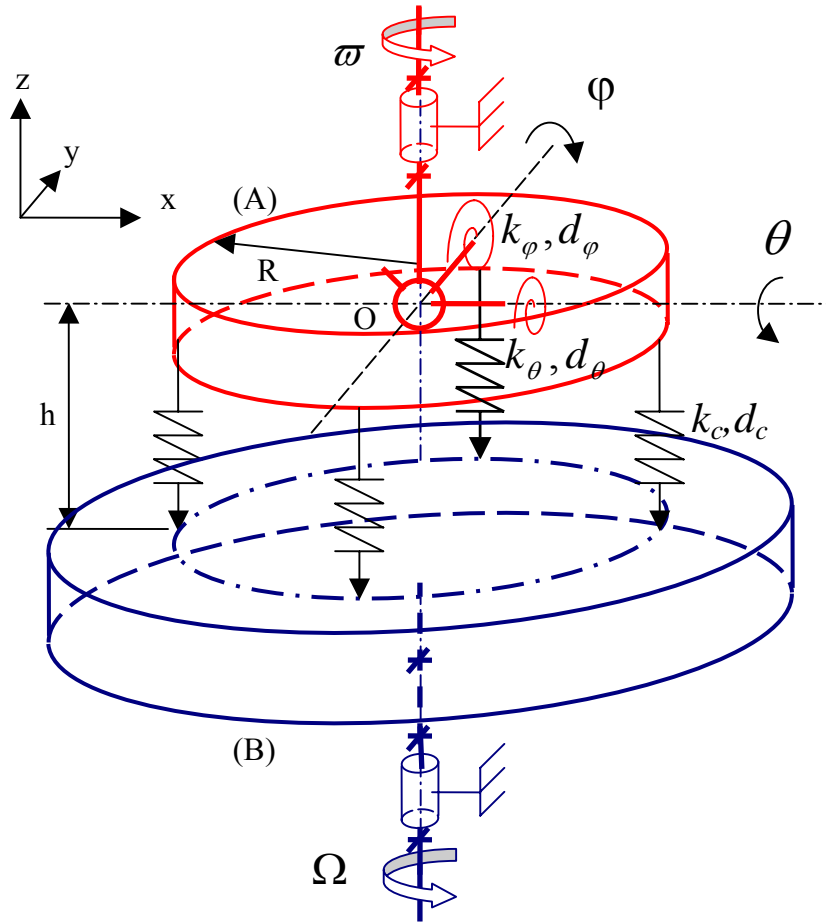

Fig. 1 Model of mode-coupling between swinging modes of a rotating disc with friction

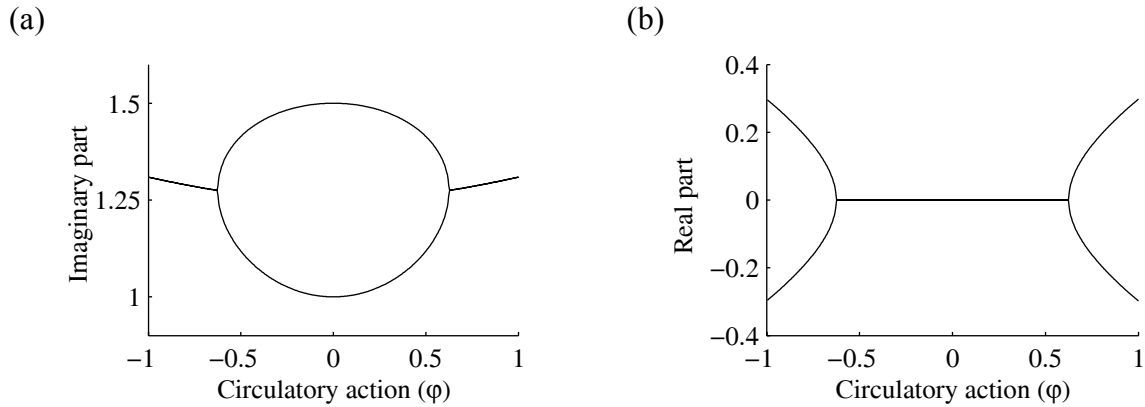

Fig. 2 Eigenvalues of an undamped purely circulatory system versus the amplitude of the circulatory action, imaginary parts (a) and real parts (b), for a natural pulsation ratio $\alpha = 1.5$

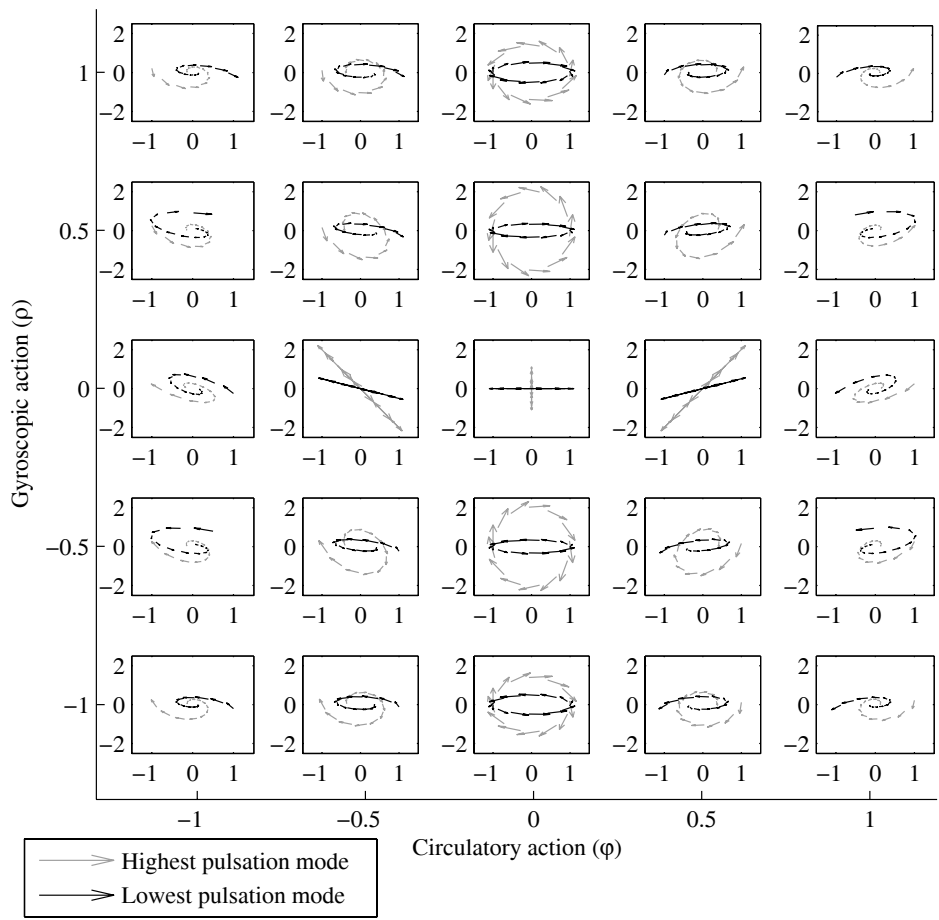

Fig. 3 Modal trajectories for one period of an undamped system in the circulatory/gyroscopic actions plane, for a natural pulsation ratio $\alpha = 1.5$; for each graph the X-coordinate is the first variable and the Y-coordinate is the second variable

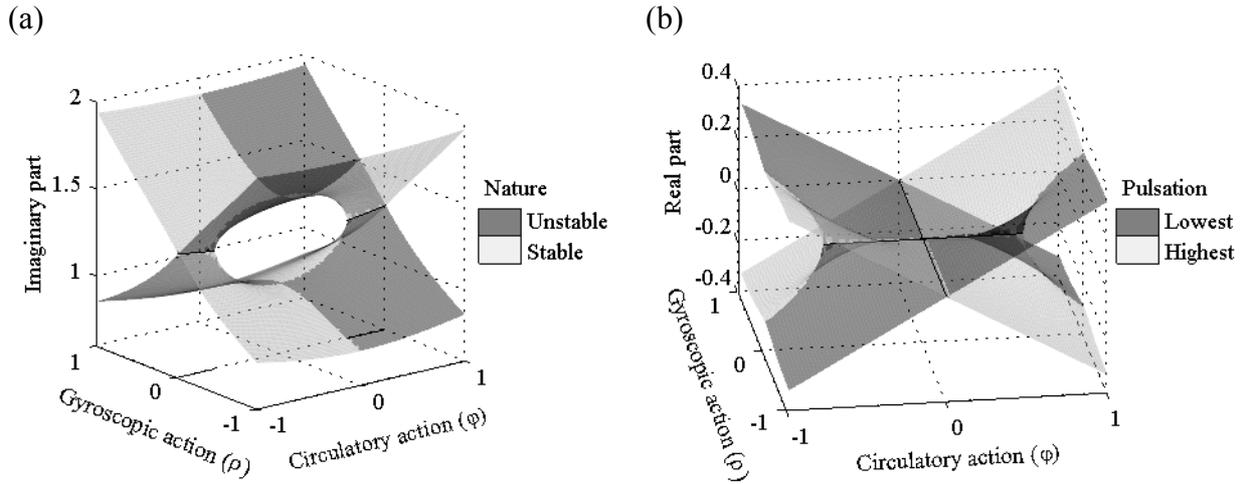

Fig. 4 Pulsations (a) and associated exponential growth rates (b) of an undamped system in the circulatory/gyroscopic actions plane, for a natural pulsation ratio $\alpha = 1.5$

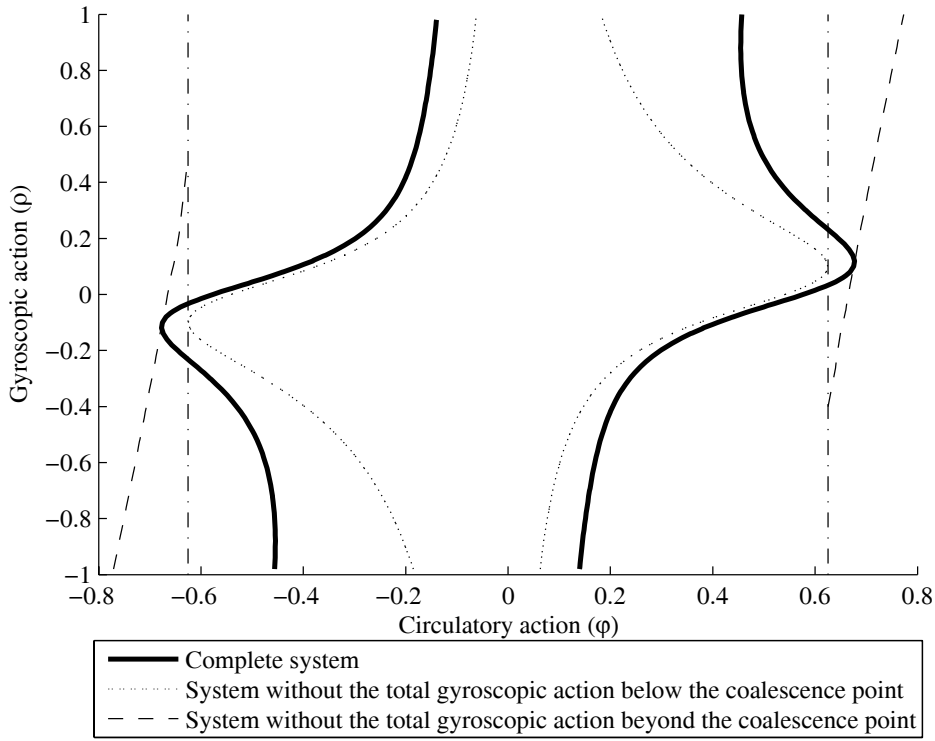

Fig. 5 Comparison of the stability frontiers in the circulatory/gyroscopic actions plane obtained for $\alpha = 1.5$, $\beta = 2$ and $\xi = 5\%$, with and without consideration for the total gyroscopic term in the eigenbase of the stiffness matrix; dashed lines denotes the coalescence points

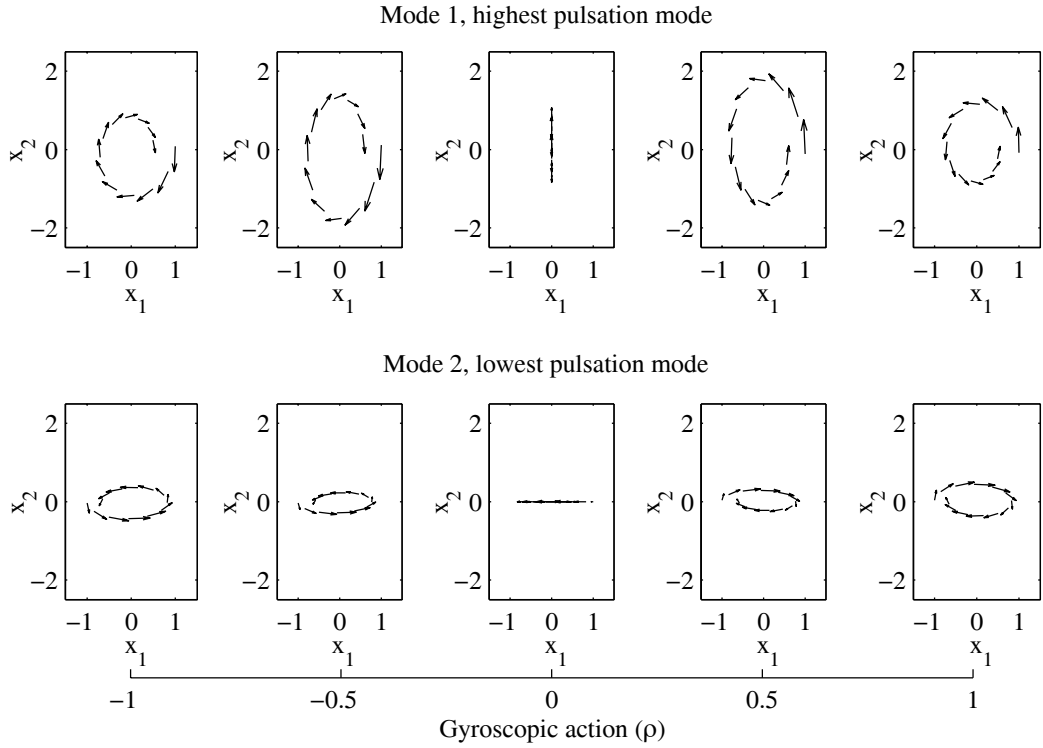

Fig. 6 Modal trajectories for one period of a damped purely gyroscopic system versus the amplitude of the gyroscopic action (ρ at -1, -0.5, 0, 0.5 and 1), for a natural pulsation ratio $\alpha = 1.5$ and a damping factor $\xi = 10\%$ and a damping ratio $\beta = 0.5$

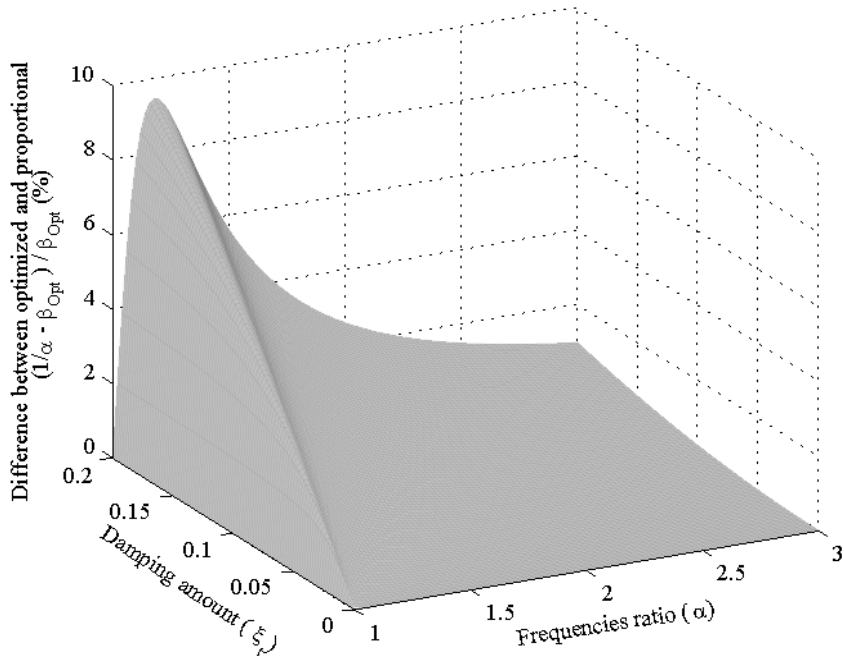

Fig. 7 Difference between the optimized damping distribution and the proportional distribution for a damped purely circulatory system in the damping amount/pulsations ratio plane

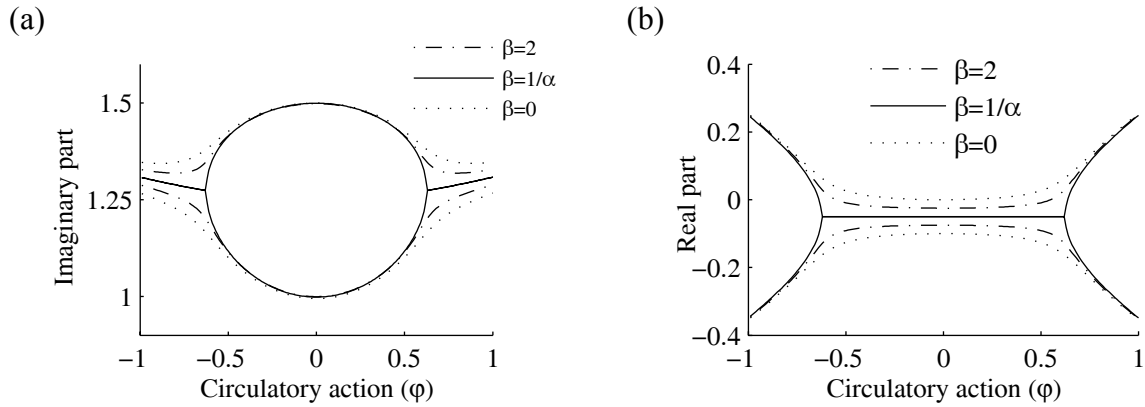

Fig. 8 Evolution of the eigenvalues of a purely circulatory system versus the amplitude of the circulatory action, imaginary parts (a) and real parts (b), for a natural pulsation ratio $\alpha = 1.5$, for a total damping amount of 10% and various damping ratio (β at 0, 1/α≈0.66 and 2)

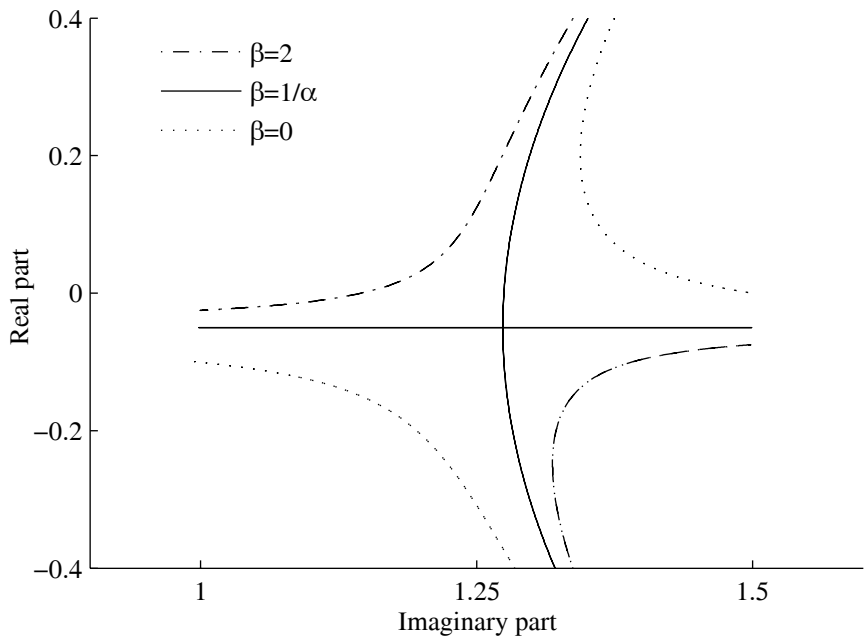

Fig. 9 Locus of the eigenvalues of a purely circulatory system in the complex plane with parametric variations of the amplitude of the circulatory action, for a natural pulsation ratio $\alpha = 1.5$, for a total damping amount of 10% and various damping ratio (β at 0, 1/α≈0.66 and 2)

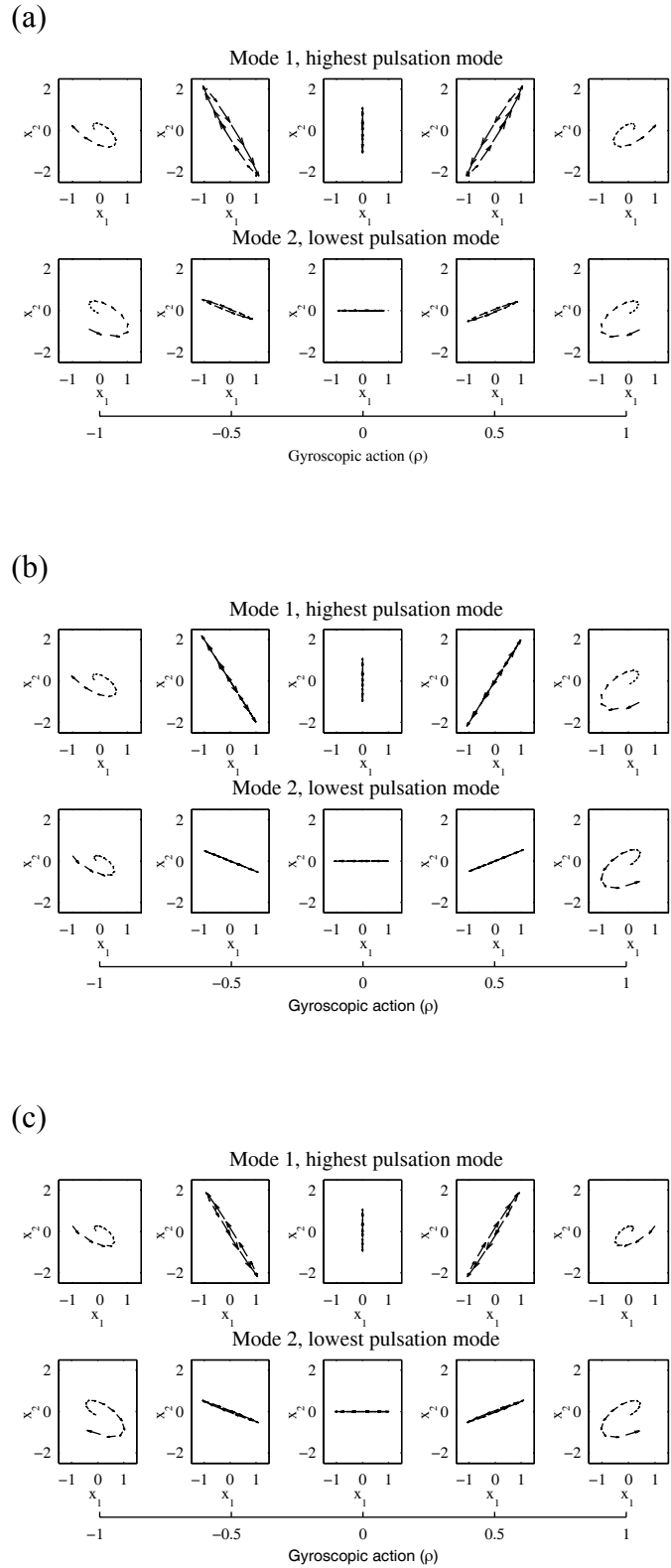

Fig. 10 Modal trajectories for one period of a damped purely circulatory system versus the amplitude of the circulatory action (φ at -1, -0.5, 0, 0.5 and 1), for a natural pulsation ratio $\alpha = 1.5$, for a total damping amount of 10% and various damping ratio β=0 ( a), β=1/α≈0.66 (b) and β=2 (c)

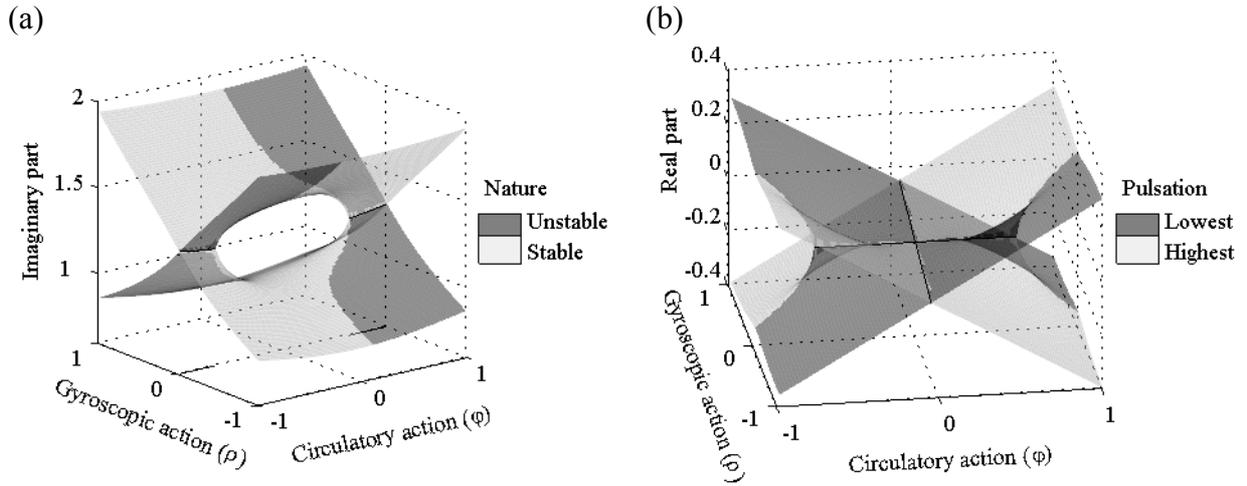

Fig. 11 Pulsations (a) and associated exponential growth rates (b) of a proportionally damped system in the circulatory/gyroscopic actions plane, for a natural pulsation ratio $\alpha = 1.5$ and a total damping amount of 10%

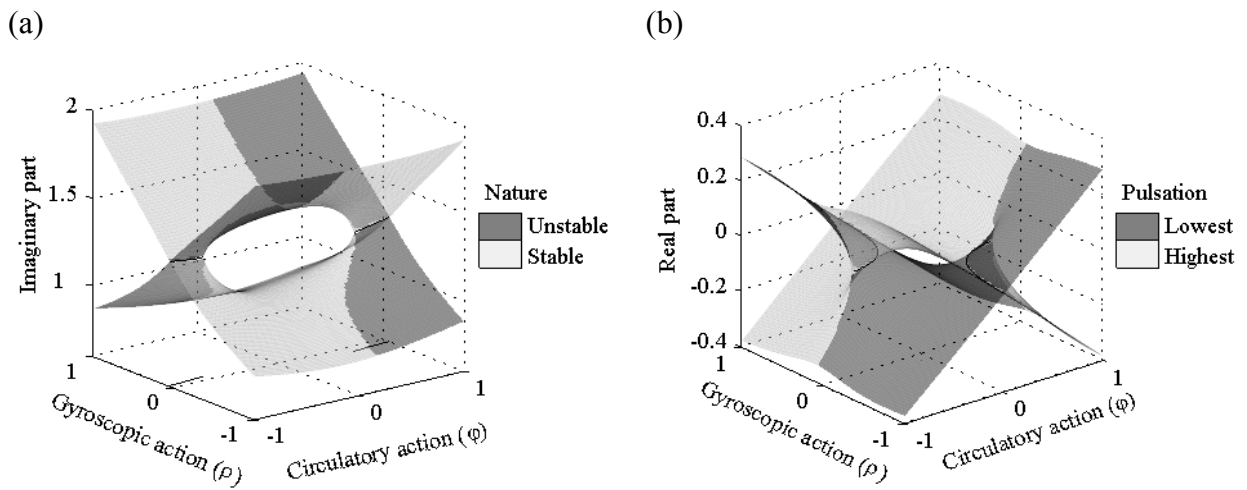

Fig. 12 Pulsations (a) and associated exponential growth rates (b) of an unbalancedly damped system in the circulatory/gyroscopic actions plane, for a natural pulsation ratio $\alpha = 1.5$ and a damping ratio of 10% on the first DOF ($\beta = 0$)

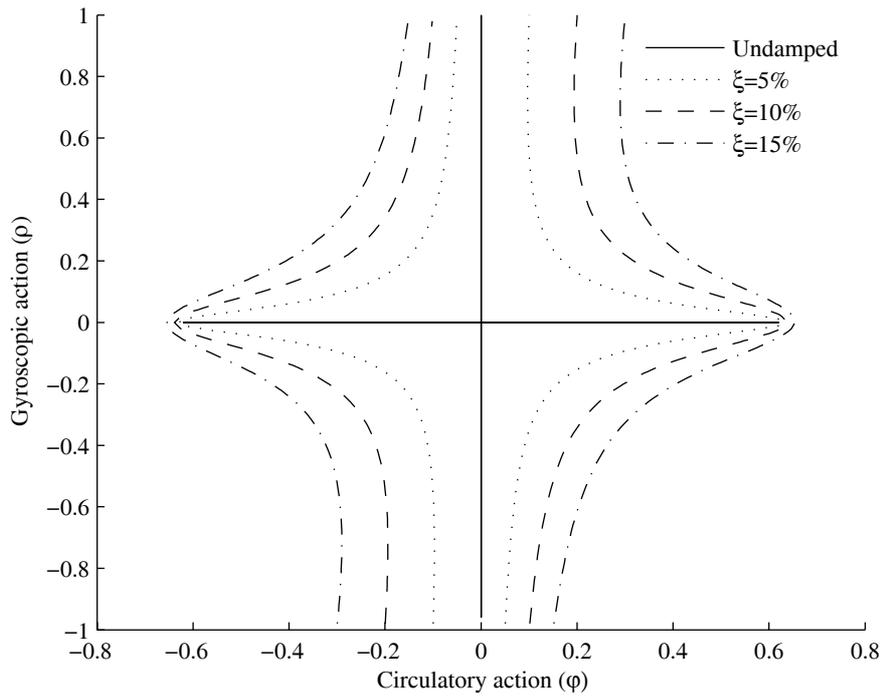

Fig. 13 Stability frontiers in the circulatory/gyroscopic actions plane obtained for $\alpha = 1.5$, and various values of the total damping amount in the proportional damping configuration

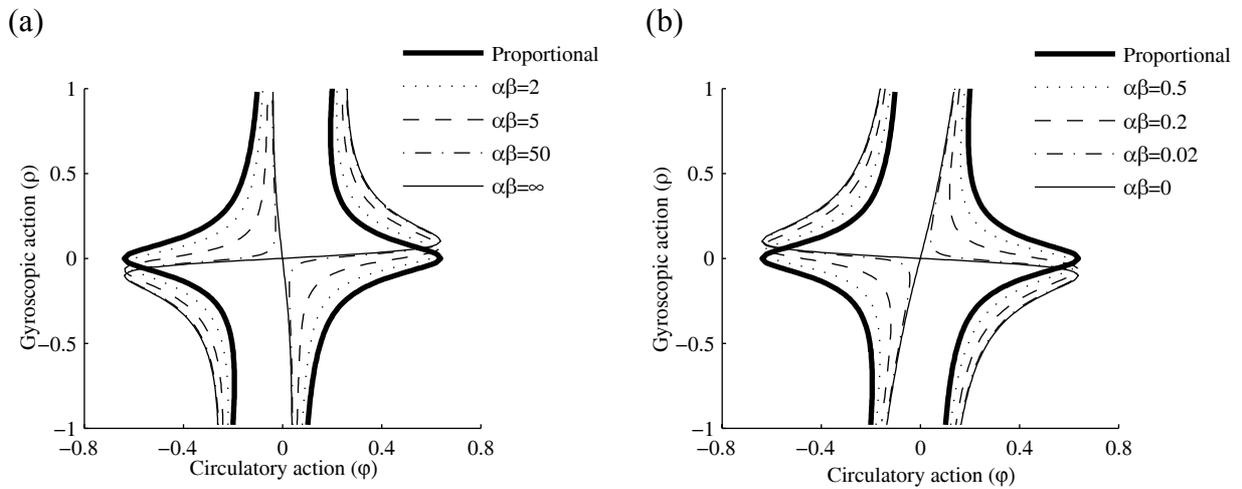

Fig. 14 Stability frontiers in the circulatory/gyroscopic actions plane obtained for $\alpha = 1.5$, for a total damping amount of 10% and various damping distributions, the 2$^{nd}$ DOF being overdamped (a), or underdamped (b) relatively to the proportional configuration

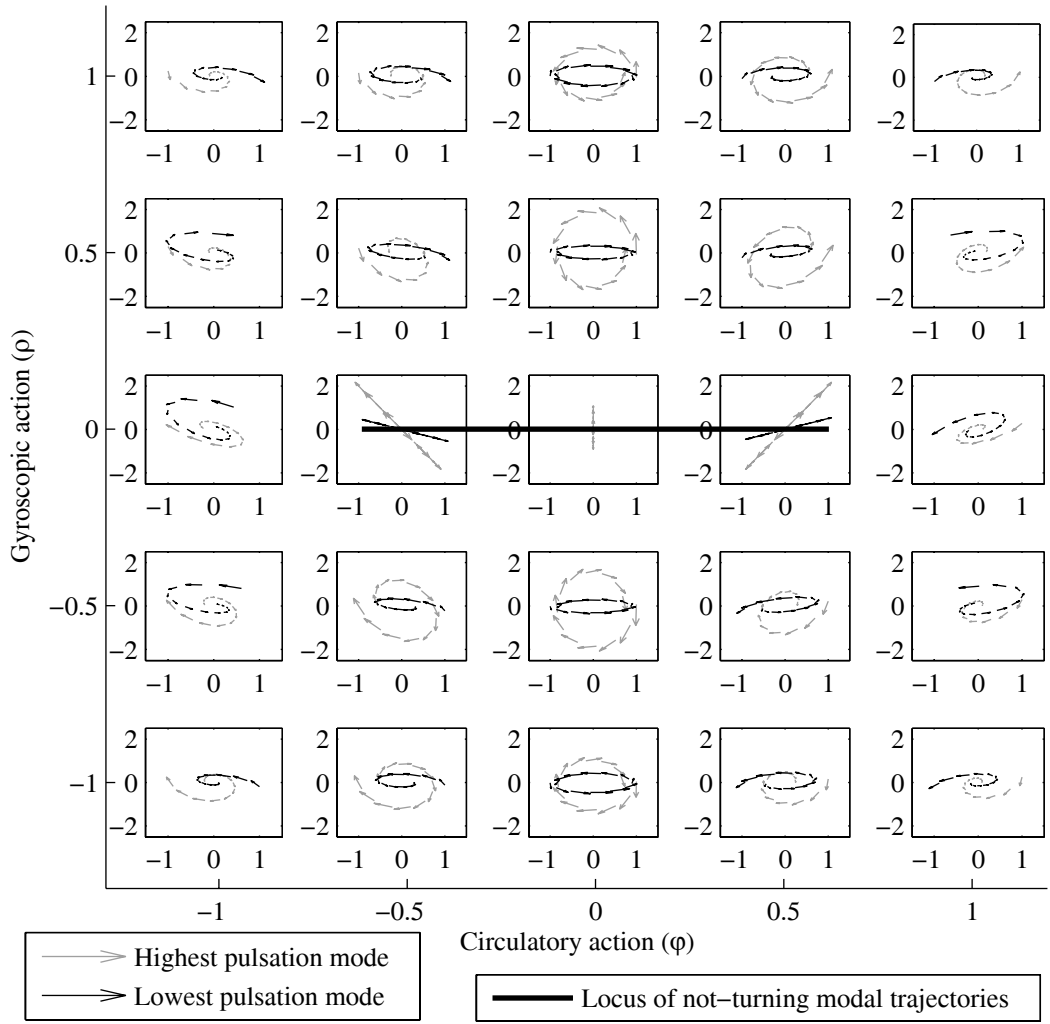

Fig. 15 Locus of not-turning modal trajectories and modal trajectories for one period of an proportionally damped system in the circulatory/gyroscopic actions plane, for a natural pulsation ratio $\alpha = 1.5$, and a total damping amount of 10%; for each graph the X-coordinate is the first variable and the Y-coordinate is the second variable

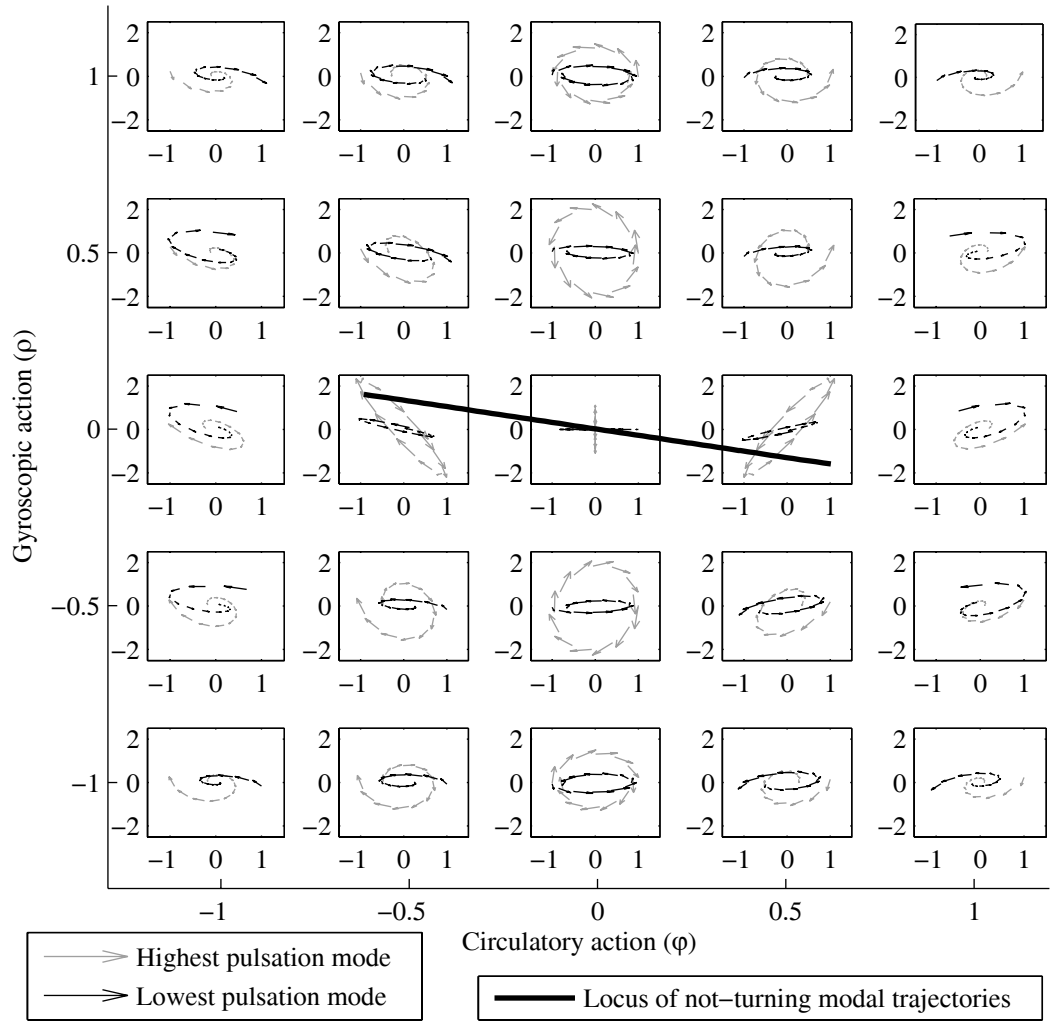

Fig. 16 Locus of not-turning modal trajectories and modal trajectories for one period of an unbalancedly damped system in the circulatory/gyroscopic actions plane, for a natural pulsation ratio $\alpha = 1.5$, and a total damping amount of 10% on the first DOF ($\beta = 0$); for each graph the X-coordinate is the first variable and the Y-coordinate is the second variable